\documentclass{kluwer}

\begin{document}      

\begin{article}                                                     
\begin{opening}         
\title{SOME HOMOGENEOUS BIANCHI TYPE IX VISCOUS FLUID COSMOLOGICAL MODELS WITH A 
VARYING $\Lambda$ }
\author{ANIRUDH PRADHAN\thanks{Corresponding Author}}
\runningtitle{SOME HOMOGENEOUS  BIANCHI TYPE IX  COSMOLOGICAL MODELS ....}
\runningauthor{A. PPRADHAN, S. K. SRIVASTAV AND M. K. YADAV }
\author{SUDHIR KUMAR SRIVASTAV}
\institute{Department of Mathematics, Hindu Post-graduate College,
Zamania-232 331, Ghazipur, U. P., India; \\
E-mail:acpradhan@yahoo.com, pradhan@iucaa.ernet.in}
\author{MAHESH KUMAR YADAV}
\institute{Department of Mathematics, Stani Memorial College (IIRM Campus),
Mansarovar, Jaipur-302 020, India} 

\date{\today}

\begin{abstract} 
Some  Bianchi type IX viscous fluid cosmological models  are investigated. To get a
solution a supplementary condition between metric potentials is used. The viscosity
coefficient of bulk viscous fluid is assumed to be a power function of mass density 
whereas the coefficient of shear viscosity is considered as proportional to scale 
of expansion in the model. The cosmological constant $\Lambda$ is found to be positive
and is a decreasing function of time which is supported by results from recent supernova
observations. Some physical and geometric properties of the models are also discussed.
\end{abstract}
\end{opening}

\section{Introduction}
\vspace*{-0.5pt}
\noindent
Bianchi type IX cosmological models are important and interesting in the sense
that these models allow not only expansion but also rotation and shear and in
general are anisotropic. Bianchi type IX universes include closed FRW models. 
The homogeneous and isotropic FRW cosmological models which are used to describe 
standard cosmological models, are particular case of Bianchi type I, V, and IX 
space-times according as the constant curvature of the physical three-space, $t$ = 
constant, is zero, negative or positive. In these models, neutrino viscosity 
explains the large radiation entropy in the universe and the degree of isotropy of
the cosmic background radiation. The standard cosmological models are too 
restrictive because of the insistence on the isotropy of the physical three-space 
and several attempts have been made to study non-standard cosmological models e. g.
MacCallum (1979); Narlikar and Kembhavi (1980); and Narlikar 1983). It is therefore
interesting to carry out the detailed studies of gravitational fields which are 
described by space-time of various Bianchi types. Vaidya and Patel (1986) have studied 
spatially homogeneous space-time of Bianchi type IX and they have outlined general
scheme for the derivation of exact solutions of Einstein's field equations in presence
of perfect fluid and pure radiation fields. Krori {\it te al.} (1990); Chakraborty 
and Nandy (1992) have investigated cosmological models of Bianchi type II, VIII and IX.
There are many other researchers (Uggla and Zur-Muhlem, 1990; Burd, Buric Ellis, 1990;
King, 1991; Paternoga and Graham, 1996), who have studied Bianchi type IX space-time in
different context.  Recently Bali and Dave (2001, 2003) have investigated Bianchi type 
IX string cosmological models. \\

Models with a dynamic cosmological term $\Lambda (t)$ are becoming popular as they solve
the cosmological constant problem in a natural way. There is significant observational 
evidence for the detection of Einstein's cosmological constant, $\Lambda$ or a component 
of material content of the universe that varies slowly with time and space and so acts 
like $\Lambda$. Recent cosmological observations by High -z Supernova Team and Supernova 
Cosmological Project (Garnavich {\it et al.}, 1998; Perlmutter {\it et al.}, 1997, 1998, 
1999; Riess {\it et al.}, 1998; Schmidt {\it et al.}, 1998) suggest the existence of a 
positive cosmological constant $\Lambda$ with magnitude 
$\Lambda (G \hbar /c^{3}) \approx 10^{-123}$. These observations on magnitude and red-shift
of type Ia supernova suggest that our universe may be a accelerating with a large function
of the cosmological density in the form of the cosmological $\Lambda$-term. Earlier researchers 
on this topic, are contained in Zeldovich (1968), Weinberg (1972), Dolgov (1983, 1990), 
Bertolami (1986), Ratra and Peebles (1988), Carroll, Press and Turner (1992). Some
of the recent discussions on the cosmological constant ``problem'' and consequence on
cosmology with a time-varying cosmological constant have been discussed by Dolgov 
(1993,1997), Tsagas and Maartens (2000), Sahni and Starobinsky (2000), Peebles (2002), 
Padmanabhan (2003), Vishwakarma (1999, 2000, 2001, 2002), and Pradhan {\it et al.} 
(2001, 2002, 2003). This motivates us to study the cosmological models in which $\Lambda$ 
varies with time. \\ 

The majority of the studies in cosmology involve a perfect fluid. However, 
observed physical phenomena such as the large entropy per baryon and the 
remarkable degree of isotropy of the cosmic microwave background radiation 
suggests analysis of dissipative effects in cosmology. Furthermore, there are
several processes which are expected to give rise to viscous effects. These
are the decoupling of neutrinos during the radiation era and the decoupling
of radiation and matter during the recombination era. Bulk viscosity is
associated with the GUT phase transition and string creation. The model 
studied by Murphy (1973) possessed an interesting feature in that the 
big bang type of singularity of infinite space-time curvature does not occur 
to be a finite past. However, the relationship assumed by Murphy between the
viscosity coefficient and the matter density is not acceptable at large 
density. The effect of bulk viscosity on the cosmological evolution has been 
investigated by a number of authors in the framework of general theory of 
relativity (Padmanabhan and Chitre, 1987; Johri and Sudarshan, 1988; Maartens, 
1995; Zimdahl, 1996; Pradhan, Sarayakar and Beesham, 1997; Kalyani and Singh 
(1997; Singh, Beesham and Mbokazi, 1998; Pradhan et al., 2001, 2002). This 
motivates to study cosmological bulk viscous fluid model.\\

Recently Bali and Yadav (2002) has investigated Bianchi type IX viscous fluid cosmological
models. Motivated by the situations discussed above, in this paper, we shall focus
upon the problem with varying cosmological constant in presence of bulk and
shear viscous fluid in an expanding universe. We do this by extending the work
of Bali and Yadav (2002) by including varying cosmological constant and the coefficient
of bulk viscosity as function of time. This paper is organized as follows. The
metric and the field equations are presented in section 2. In section 3, we
deal with the solution of the field equations in presence of viscous fluid.
Section 4 includes the solution of some particular models whereas in section 5, we deal 
with some special models. In section 6, we have given the concluding remarks.    \\   
\section{The metric and field  equations}
We consider the Bianchi type IX metric in the form
\[
ds^{2} = - dt^{2} + A^{2} dx^{2} + B^{2}dy^{2} + (B^{2} ~ \sin^{2}y + 
A^{2} ~ \cos^{2}y) ~ dz^{2}
\]
\begin{equation}
\label{eq1}
- 2 A^{2} ~ \cos y ~ dx ~  dz,
\end{equation}
where A and B are functions of $t$ only.\\
The Einstein's field equations (in gravitational units $c = 1$, $G = 1$) read as
\begin{equation}
\label{eq2}
R^{j}_{i} - \frac{1}{2} R g^{j}_{i} + \Lambda g^{j}_{i} = - 8\pi T^{j}_{i}
\end{equation}
where $R^{j}_{i}$ is the Ricci tensor; $R$ = $g^{ij} R_{ij}$ is the
Ricci scalar; and $T^{j}_{i}$ is the stress energy-tensor in the presence
of bulk stress given by Landau and Lifshitz (1963) 
\[
T^{j}_{i} = (\rho + p)v_{i}v^{j} + p g^{j}_{i} - \eta \left ({v^{j}_{i ;}} + {v^j}_{;i} 
+ v^{j} v^{l} v_{i;l} + v_{i} v^{l} {v^j}_{;l}\right) 
\]
\begin{equation}
\label{eq3}
- \left (\xi - \frac{2}{3} \eta \right)
 ~ \theta ({g^j}_{i} + v_{i} v^{j}).
\end{equation}
Here $\rho$, $p$, $\eta$ and $\xi$ are the energy density,
isotropic pressure, coefficient of shear viscosity and bulk viscous 
coefficient respectively and $v_{i}$ is the flow vector satisfying 
the relations
\begin{equation}
\label{eq4}
g_{ij} v^{i}v^{j} = - 1.
\end{equation}
The semicolon $(;)$ indicates covariant differentiation. We choose the coordinates
to be comoving, so that 
\begin{equation}
\label{eq5}
v^1 = 0 = v^2 = v^3, v^4 = 1
\end{equation}
The Einstein's field equations (\ref{eq2}) for the line element (\ref{eq1})
has been set up as
\begin{equation}
\label{eq6}
- 8 \pi \left[p - 2\eta \frac{A_{4}}{A} - \left (\xi - \frac{2}{3}\eta \right) \theta\right] =
\frac{2B_{44}}{B} + \frac{{B_{4}}^2}{B^2} + \frac{1}{B^{2}} - \frac{3A^{2}}{4 B^{4}} + \Lambda, 
\end{equation}
\begin{equation}
\label{eq7}
- 8 \pi \left[p - 2\eta \frac{B_{4}}{B} - \left (\xi - \frac{2}{3}\eta \right) \theta\right] =
\frac{A_{44}}{A} + \frac{A_{4} B_{4}}{AB} + \frac{B_{44}}{B} + \frac{A^{2}}{4 B^{4}} + \Lambda,
\end{equation}
\begin{equation}
\label{eq8}
8 \pi \rho = \frac{2 A_{4} B_{4}}{A B} +  \frac{{B_{4}}^2}{B^2} + \frac{1}{B^{2}} 
- \frac{A^{2}}{4 B^{4}} + \Lambda, 
\end{equation}
where the suffix $4$ at the symbols $A$ and $B$ denotes ordinary 
differentiation with respect to $t$ and $\theta$ is the scalar of expansion given by
\begin{equation}
\label{eq9}
\theta = {v^{i}}_{;i}.
\end{equation}
\section{Solution of the field equations}
Equations (\ref{eq6}) - (\ref{eq8}) are three  independent 
equations in seven unknowns $A$, $B$, $\rho$, $p$,  $ \eta$, $\xi$ and $\Lambda$.
For the complete determinacy of the system, we need four  extra conditions.\\
Firstly we assume a relation in metric potential as
\begin{equation}
\label{eq10}
A = B^{m}
\end{equation}
and secondly we assume that the coefficient of shear viscosity is proportional to
the scale of expansion, i. e.,
\begin{equation}
\label{eq11}
\eta  \propto  \theta
\end{equation}
where $m$ is a real number.\\
Eqs. (\ref{eq6}) and (\ref{eq7}) lead to
\begin{equation}
\label{eq12}
\frac{B_{44}}{B} + \frac{{B_{4}}^2}{B^{2}} - \frac{A_{44}}{A} - \frac{A_{4}B_{4}}{A B} - 
\frac{A^{2}}{B^{4}} + \frac{1}{B^{2}} = 16 \pi \eta \left( \frac{A_{4}}{A} - \frac{B_{4}}{B}\right).
\end{equation}
Condition (\ref{eq11}) leads to
\begin{equation}
\label{eq13}
\eta = \ell \left(\frac{A_{4}}{A} + \frac{2 B_{4}}{B}\right),
\end{equation}
where $\ell$ is a proportionality constant. \\
Equations (\ref{eq12}) together with (\ref{eq10}) and (\ref{eq13}) leads to
\begin{equation}
\label{eq14}
B B_{44} + \alpha {B_{4}}^2 = \frac{B^{2(m-1)}}{(1-m)} - \frac{1}{(1 - m)},  ~ ~ m \neq 1, 
\end{equation}
which can be rewritten as
\begin{equation}
\label{eq15}
\frac{d}{dB} (f^{2}) + \frac{2 \alpha}{B} (f^{2}) = \frac{2B^{2m - 3}}{(1- m)} - \frac{2}{(1- m)B}
\end{equation}
where 
\begin{equation}
\label{eq16}
\alpha = (1 + m) - 16 \pi \ell ~\frac{(m^{2} + m - 2)}{(1 - m)} 
\end{equation}
and
\begin{equation}
\label{eq17}
B_{4} = f(B)
\end{equation}
From (\ref{eq15}), we obtain
\begin{equation}
\label{eq18}
\left(\frac{dB}{dt}\right)^{2} = \left[\frac{B^{2(m-1)}}{(1 - m)(m + \alpha - 1)} + 
\frac{\beta}{B^{2\alpha}} - \frac{1}{\alpha(1 - m)}\right],
\end{equation}
where $\beta$ is a constant of integration. 
After a suitable transformation of coordinates, the metric (\ref{eq1}) reduces to the form
\[
ds^{2} = - \left[\frac{T^{2(m - 1)}}{(1 - m) ~ (m + \alpha - 1)} + \frac{\beta}{T^{2\alpha}}
- \frac{1}{\alpha(1 - m)}\right]^{-1} ~ dT^{2} + T^{2m} ~ dx^{2} +
\]
\begin{equation}
\label{eq19}
T^{2} ~ dy^{2} + (T^{2} ~ \sin^{2} y
+ T^{2m} ~ \cos^{2} y) ~ dz^{2} - 2 T^{2m} ~ \cos y ~ dx ~dy, 
\end{equation}
where $B = T$.\\
The pressure and density for the model (\ref{eq19}) are given by
\[
8 \pi p = K_{1} ~ T^{2(m -2)} + \frac{K_{2}}{T^{2}} + \frac{\beta K_{3}}
{T^{2(\alpha + 1)}} + 
\]
\begin{equation}
\label{eq20} 
8\pi \xi (m + 2)\sqrt{\left[\frac{T^{2(m - 2)}}{(m + \alpha - 1)(1 - m)} 
+ \frac{\beta}{T^{2(\alpha + 1)}} - \frac{1}{\alpha (1 - m) T^{2}}\right]} - \Lambda,
\end{equation}
\begin{equation}
\label{eq21}
8 \pi \rho = K_{4} ~ T^{2(m - 2)} + \frac{\beta (2m + 1)}{T^{2(\alpha + 1)}} + 
\frac{K_{5}}{T^{2}} + \Lambda, 
\end{equation}
where \\
$
K_{1} = \frac{(-m^{2})(64 \pi \ell + 21) - m(64 \pi \ell - 3\alpha + 6) + 
(128 \pi \ell - 3 \alpha + 15)}{12(m + \alpha - 1)(1 - m)}, \\
K_{2} = \frac{m^{2}(16 \pi \ell + 3) + m(16 \pi \ell) - 32 \pi \ell}{3\alpha (1 - m)}, \\
K_{3} = \frac{1}{3}[(-m^{2})(16 \pi \ell + 3) - m(16 \pi \ell - 3 \alpha) + 
(32 \pi \ell + 3\alpha)], \\
K_{4} = \frac{m^{2} + m(\alpha + 6) -(\alpha - 5)}{4(m + \alpha - 1)(1 - m)}, \\
K_{5} = \frac{(-m)(\alpha + 2) + (\alpha - 1)}{\alpha(1 - m)}. \\
$
For the specification of $\xi$, we assume that the fluid obeys an equation of state of the
form
\begin{equation}
\label{eq22}
p = \gamma \rho,
\end{equation}
where $\gamma (0 \leq \gamma \leq 1)$ is constant. \\
Thus,  given $\xi(t)$ we can solve for the cosmological parameters. In most of the 
investigation involving bulk viscosity is assumed to be a simple power function of 
the energy density (Pavon, 1991; Maartens, 1995; Zimdahl, 1996) 
\begin{equation}
\label{eq23}
\xi(t) = \xi_{0} \rho^{n},
\end{equation}
where $\xi_{0}$ and $n$ are constants. If $n = 1$, Equation (\ref{eq23}) may correspond 
to a radiative fluid (Weinberg, 1972). However, more realistic models (Santos, 1985) are 
based on $n$ lying in the regime $0 \leq n \leq \frac{1}{2}$. \\
On using (\ref{eq23}) in (\ref{eq20}), we obtain
\[
8 \pi p = K_{1} ~ T^{2(m -2)} + \frac{K_{2}}{T^{2}} + \frac{\beta K_{3}}
{T^{2(\alpha + 1)}} + 
\]
\begin{equation}
\label{eq24} 
8\pi \xi_{0}~ \rho^{n}~ (m + 2)\sqrt{\left[\frac{T^{2(m - 2)}}{(m + \alpha - 1)(1 - m)} 
+ \frac{\beta}{T^{2(\alpha + 1)}} - \frac{1}{\alpha (1 - m) T^{2}}\right]} - \Lambda.
\end{equation}
\subsection{Model I:~ ~ ~ ~ Solution for $\xi = \xi_{0}$}
When $n = 0$, Equation (\ref{eq23}) reduces to $\xi = \xi_{0}$ = constant. Hence in 
this case Equation (\ref{eq24}), with the use of (\ref{eq21}) and (\ref{eq22}), leads to
\[ 
8 \pi (1 + \gamma) \rho = (K_{1} + K_{4}) ~ T^{2(m - 2)} + \frac{(K_{2} + K_{5})}{T^{2}}
+ \frac{\beta (K_{3} + 2m + 1)}{T^{2(\alpha + 1)}} 
\]
\begin{equation}
\label{eq25}
+ 8\pi \xi_{0}  (m + 2)\sqrt{\left[\frac{T^{2(m - 2)}}{(m + \alpha - 1)(1 - m)} 
+ \frac{\beta}{T^{2(\alpha + 1)}} - \frac{1}{\alpha (1 - m) T^{2}}\right]}.
\end{equation}
Eliminating $\rho(t)$ between Equations (\ref{eq21}) and (\ref{eq25}), we have
\[ 
(1 + \gamma) \Lambda  = (K_{1} - \gamma  K_{4}) ~ T^{2(m - 2)} + \frac{(K_{2} -
\gamma  K_{5})}{T^{2}} + \frac{\beta \left(K_{3} - \gamma ( 2m + 1)\right)}{T^{2(\alpha + 1)}} 
\]
\begin{equation}
\label{eq26}
+ 8\pi \xi_{0} (m + 2)\sqrt{\left[\frac{T^{2(m - 2)}}{(m + \alpha - 1)(1 - m)} 
+ \frac{\beta}{T^{2(\alpha + 1)}} - \frac{1}{\alpha (1 - m) T^{2}}\right]}.
\end{equation}
\subsection{Model II:~ ~ ~ ~ Solution for $\xi = \xi_{0}\rho$}
When $n = 1$, Equation (\ref{eq23}) reduces to $\xi = \xi_{0} \rho$. Hence in 
this case Equation (\ref{eq24}), with the use of (\ref{eq21}) and (\ref{eq22}), 
leads to
\[ 
8 \pi \rho = \frac {1}{\left[1 + \gamma -  \xi_{0} (m + 2)\sqrt{\left[\frac{T^{2(m - 2)}}
{(m + \alpha - 1)(1 - m)} + \frac{\beta}{T^{2(\alpha + 1)}} - \frac{1}{\alpha (1 - m)
 T^{2}}\right]}\right]} \times
\]
\begin{equation}
\label{eq27}
\left[K_{1} ~ T^{2(m - 2)} + \frac{K_{2}}{T^{2}} + \frac{\beta ~ K_{3}}
{T^{2(\alpha + 1)}}\right].
\end{equation}
Eliminating $\rho(t)$ between Equations (\ref{eq21}) and (\ref{eq27}), we have
\[ 
\Lambda = \frac {1}{\left[1 + \gamma -  \xi_{0} (m + 2)\sqrt{\left[\frac{T^{2(m - 2)}}
{(m + \alpha - 1)(1 - m)} + \frac{\beta}{T^{2(\alpha + 1)}} - \frac{1}{\alpha (1 - m)
 T^{2}}\right]}\right]} \times
\]
\begin{equation}
\label{eq28}
\left[K_{1} ~ T^{2(m - 2)} + \frac{K_{2}}{T^{2}} + \frac{\beta ~ K_{3}}
{T^{2(\alpha + 1)}}\right] - \left(K_{4}T^{2(m - 2)} + \frac{K_{5}}{T^{2}} + 
\frac{\beta (2m + 1)}{T^{2(\alpha + 1)}}\right).
\end{equation}
From Equations (\ref{eq26}) and (\ref{eq28}), we observe that when $\alpha > 0$
 and $m < 2$, the positive cosmological constant is a decreasing function of time and  
approaches a small value in the present epoch. \\
{\bf Some Physical Aspects of the Models}: \\
With regard to the kinematical properties of the velocity vector $v^{i}$ in the metric 
(\ref{eq19}), a straight forward calculation leads to the following expressions for the
scalar of expansion $(\theta)$ and for the shear $(\sigma)$ of the fluid.
\begin{equation}
\label{eq29}
\theta = (m + 2)\left[\sqrt{\left[\frac{T^{2(m - 2)}}{(m + \alpha - 1)(1 - m)} 
+ \frac{\beta}{T^{2(\alpha + 1)}} - \frac{1}{\alpha (1 - m) T^{2}}\right]}\right]
\end{equation}
\begin{equation}
\label{eq30}
\sigma  = \sqrt{\frac{2}{3}} ~ (1 - m)\left[\sqrt{\left[\frac{T^{2(m - 2)}}{(m + \alpha - 1)
(1 - m)} + \frac{\beta}{T^{2(\alpha + 1)}} - \frac{1}{\alpha (1 - m) T^{2}}\right]}\right]
\end{equation}
For $\alpha > 0$ and $m < 2$, the expansion factor $\theta$ is a decreasing function of $T$
and approaches, asymptotically to zero with $\rho$ and $p$ also approaching to zero as 
$ T \rightarrow \infty$.\\
\section{Particular Models}
If we set $ m = 2$, then the geometry of the space-time (\ref{eq19}) reduces to the
form
\[
- \left[\frac{\beta}{T^{2(3 + 64 \pi \ell)}} - \frac{T^{2}}{4(1 + 16 \pi \ell)} + \frac{1}
{(3 + 64 \pi \ell)}\right] ~ dT^{2}
\]
\begin{equation}
\label{eq31}
+ ~ T^{4} ~ dx^{2} + T^{2} ~ dy^{2} + (T^{2}~ \sin^{2} y + T^{4}~ \cos^{2} y) dz^{2}
- 2 T^{4} ~ \cos y ~ dx ~ dz,
\end{equation}
where $\beta$ is an integrating constant.\\
The pressure and density of the model (\ref{eq31}) are given by
\[
8 \pi p = \frac{\beta(15 + 512 \pi \ell)}{3 T^{8(1 + 16 \pi \ell)}} - \frac{4(3 + 16 \pi \ell)}
{3(3 + 64 \pi \ell) ~ T^{2}} + \frac{(9 + 8 \pi \ell)}{6(1 + 16 \pi \ell)}
\]
\begin{equation}
\label{eq32}
+ ~ (32 \pi \xi) ~ \sqrt{\left[\frac{\beta}{T^{8(1 + 16 \pi \ell)}} + \frac{1}{(3 + 64 \pi \ell) ~ T^{2}}
- \frac{1}{4(1 + 16 \pi \ell)}\right]} - \Lambda
\end{equation}
\begin{equation}
\label{eq33}
8 \pi \rho = \frac{5\beta}{T^{8(1 + 16 \pi \ell)}} + \frac{8(1 + 8 \pi \ell)}{(3 + 64 \pi \ell)~ T^{2}}
- \frac{(3 + 8 \pi \ell)}{2(1 + 16 \pi \ell)} + \Lambda.
\end{equation}
\subsection{Model I:~ ~ ~ ~ Solution for $\xi = \xi_{0}$}
When $n = 0$, Equation (\ref{eq23}) reduces to $\xi = \xi_{0}$ = constant. Hence in 
this case Equation (\ref{eq32}), with the use of (\ref{eq33}) and (\ref{eq22}), leads to
\[
8 \pi (1 + \gamma) \rho = \frac{2\beta(15 + 256 \pi \ell)}{3T^{8(1 + 16\pi \ell)}} +
~ \frac{4(15 + 32 \pi \ell)}{3(3 + 64\pi \ell)T^{2}} - \frac{16 \pi \ell}{3(1 + 16 \pi \ell)}
\]
\begin{equation}
\label{eq34}
+ ~ (32 \pi \xi_{0}) ~ \sqrt{\left[\frac{\beta}{T^{8(1 + 16 \pi \ell)}} + 
\frac{1}{(3 + 64 \pi \ell) ~ T^{2}} - \frac{1}{4(1 + 16 \pi \ell)}\right]}
\end{equation}
Eliminating $\rho(t)$ between (\ref{eq33}) and (\ref{eq34}), we obtain
\[
(1 + \gamma) \Lambda = \frac{\beta(15 + 512 \pi \ell - 15\gamma)}{3T^{8(1 + 16 \pi \ell)}}
- \frac{4\left(3 + 16 \pi \ell + 6(3 + 8\pi \ell)\gamma\right)}{3(3 + 64 \pi \ell) T^{2}}
+ \frac{\left(9 - 8 \pi \ell + 3(3 + 8 \pi \ell)\gamma\right)}{6(1 + 16 \pi \ell)}
\]
\begin{equation}
\label{eq35}
+ ~ (32 \pi \xi_{0}) ~ \sqrt{\left[\frac{\beta}{T^{8(1 + 16 \pi \ell)}} + 
\frac{1}{(3 + 64 \pi \ell) ~ T^{2}} - \frac{1}{4(1 + 16 \pi \ell)}\right]}
\end{equation}
\subsection{Model II:~ ~ ~ ~ Solution for $\xi = \xi_{0}\rho$}
When $n = 1$, Equation (\ref{eq23}) reduces to $\xi = \xi_{0} \rho$. Hence in 
this case Equation (\ref{eq32}), with the use of (\ref{eq33}) and (\ref{eq22}), 
leads to
\[
\rho = \frac{1}{8\pi \left[1 + \gamma - 4 \xi_{0} ~ \sqrt{\left[\frac{\beta}{T^{8(1 + 16 \pi \ell)}} + 
\frac{1}{(3 + 64 \pi \ell) ~ T^{2}} - \frac{1}{4(1 + 16 \pi \ell)}\right]}\right]}\times
\]
\begin{equation}
\label{eq36}
 \frac{2\beta(15 + 256 \pi \ell)}{3T^{8(1 + 16\pi \ell)}} +
~ \frac{4(15 + 32 \pi \ell)}{3(3 + 64\pi \ell)T^{2}} - \frac{16 \pi \ell}{3(1 + 16 \pi \ell)}
\end{equation}
Eliminating $\rho(t)$ between (\ref{eq33}) and (\ref{eq36}), we obtain 
\[
\Lambda = \frac{1}{\left[1 + \gamma - 4 \xi_{0} ~ \sqrt{\left[\frac{\beta}{T^{8(1 + 16 \pi \ell)}} + 
\frac{1}{(3 + 64 \pi \ell) ~ T^{2}} - \frac{1}{4(1 + 16 \pi \ell)}\right]}\right]}\times
\]
\[
\left[ \frac{2\beta (15 + 256 \pi \ell)}{3T^{8(1 + 16\pi \ell)}} +
~ \frac{4(15 + 32 \pi \ell)}{3(3 + 64\pi \ell)T^{2}} - \frac{16 \pi \ell}{3(1 + 16 \pi \ell)}\right]
 \]
\begin{equation}
\label{eq37}
- \frac{5\beta}{T^{8(1 + 16 \pi \ell)}} - \frac{8(1 + 8 \pi \ell)}{(3 + 64 \pi \ell)~ T^{2}}
+ \frac{(3 + 8 \pi \ell)}{2(1 + 16 \pi \ell)} + \Lambda.
\end{equation}
From Equations (\ref{eq35}) and (\ref{eq37}), we observe that the positive cosmological constant 
is a decreasing function of time and  approaches a small value in the present epoch. \\
{\bf Some Physical Aspects of the Models}: \\
The expansion$(\theta)$ and the shear$(\sigma)$ in the model (\ref{eq31}) are given by
\begin{equation}
\label{eq38} 
 \theta = 4 ~ \sqrt{\left[\frac{\beta}{T^{8(1 + 16 \pi \ell)}} + 
\frac{1}{(3 + 64 \pi \ell) ~ T^{2}} - \frac{1}{4(1 + 16 \pi \ell)}\right]}
\end{equation}
\begin{equation}
\label{eq39}
\sigma =  \sqrt{\frac{2}{3}\left[\frac{\beta}{T^{8(1 + 16 \pi \ell)}} + 
\frac{1}{(3 + 64 \pi \ell) ~ T^{2}} - \frac{1}{4(1 + 16 \pi \ell)}\right]}
\end{equation}
The expansion factor $\theta$ in the model is a decreasing function of $T$. Since 
$lim_{T \rightarrow \infty} \frac{\sigma}{\theta} \neq 0$, hence the models do not 
approach isotropy for large values of $T$. The model have singularity at $T = 0$ 
which is real physical singularity.
\section{Special  Models}
If we set $ m = 2$ and $\ell = - \frac{1}{32\pi}$, Equation (\ref{eq18}) leads to
\begin{equation}
\label{eq40}
\frac{\sqrt{2} ~ B ~ dB}{\sqrt{(2B^{2} - B^{4} + 2\beta)}} = dt,
\end{equation}
which on integration gives
\begin{equation}
\label{eq41}
B^{2} = 1 + M ~ \sin(\sqrt{2} ~ t + 2 N),
\end{equation}
where
$
M = \sqrt{2 \beta +1}
$
and $N$ ia a constant of integration. Hence, we obtain 
\begin{equation}
\label{eq42}
A = B^{2} =  1 + M ~ \sin(\sqrt{2} ~ t + 2 N),
\end{equation}
Using the transformations \\
\[
\sqrt{2} ~ t + 2 N = T, 
\]
\[
x = X,
\]
\[
y = Y, 
\]
\begin{equation}
\label{eq43}
z = Z,
\end{equation}
the metric (\ref{eq1}) takes the form
\[
ds^{2} = - \frac{dT^{2}}{2} + ( 1 + M ~ \sin T)^{2} ~ dX^{2} + (1 + M ~ \sin T)dY^{2} +
\]
\[
[(1 + M ~ \sin T) ~ \sin^{2}Y + (1 + M ~ \sin T)^{2}~ \cos ^{2}Y]~dZ^{2}
\]
\begin{equation}
\label{eq44}
- 2(1 + M ~ \sin T)^{2} ~ \cos Y ~ dX ~ dZ.
\end{equation}
The pressure and density for the model  (\ref{eq44}) are given by
\begin{equation}
\label{eq45}
8 \pi p = \frac{16\sqrt{2} ~ \pi \xi M \cos T}{(1 + M \sin T)} + \frac{[35 M^{2} \sin^{2} T
+ 30 M \sin T - 2 M^{2} -3]}{12(1 + M \sin T)^{2}} - \Lambda,
\end{equation}
\begin{equation}
\label{eq46}
8 \pi \rho = \frac{[3 + 10 M^{2} + 2 M \sin T - 11 M^{2} \sin^{2} T]}{4(1 + M \sin T)^{2}} + \Lambda.
\end{equation}
\subsection{Model I:~ ~ ~ ~ Solution for $\xi = \xi_{0}$}
When $n = 0$, Equation (\ref{eq23}) reduces to $\xi = \xi_{0}$ = constant. Hence in 
this case Equation (\ref{eq45}), with the use of (\ref{eq46}) and (\ref{eq22}), leads to
\begin{equation}
\label{eq47}
8 \pi (1 + \gamma) \rho = \frac{16\sqrt{2} ~ \pi \xi_{0} M \cos T}{(1 + M \sin T)} + 
\frac{[ M^{2} \sin^{2} T + 18 M \sin T + 14 M^{2} + 3]}{6(1 + M \sin T)^{2}}.
\end{equation}
Eliminating $\rho(t)$ between (\ref{eq46}) and (\ref{eq47}), we obtain 
\[
(1 + \gamma) \Lambda = \frac{16\sqrt{2} ~ \pi \xi_{0} M \cos T}{(1 + M \sin T)} + 
\frac{[ 35 M^{2} \sin^{2} T + 30 M \sin T - 2 M^{2} - 3]}{12 (1 + M \sin T)^{2}}
\]
\begin{equation}
\label{eq48}
- ~ \frac{[3 + 10 M^{2} + 2 M \sin T - 11 M^{2} \sin^{2} T]\gamma}{4(1 + M \sin T)^{2}}.
\end{equation}
\subsection{Model II:~ ~ ~ ~ Solution for $\xi = \xi_{0}\rho$}
When $n = 1$, Equation (\ref{eq23}) reduces to $\xi = \xi_{0} \rho$. Hence in 
this case Equation (\ref{eq45}), with the use of (\ref{eq46}) and (\ref{eq22}), 
leads to
\[
8 \pi \rho = \frac{1}{\left[ 1 + \gamma - \frac{2\sqrt{2} ~  \xi_{0} M \cos T}
{(1 + M \sin T)}\right]}\times
\]
\begin{equation}
\label{eq49}
\frac{[ M^{2} \sin^{2} T + 18 M \sin T + 14 M^{2} + 3]}{6(1 + M \sin T)^{2}}.
\end{equation}
Eliminating $\rho(t)$ between (\ref{eq46}) and (\ref{eq49}), we obtain 
\[
8 \pi \rho = \frac{1}{\left[ 1 + \gamma - \frac{2\sqrt{2} ~  \xi_{0} M \cos T}
{(1 + M \sin T)}\right]}\times
\]
\[
\frac{[ M^{2} \sin^{2} T + 18 M \sin T + 14 M^{2} + 3]}{6(1 + M \sin T)^{2}}
\]
\begin{equation}
\label{eq50}
- ~ \frac{[3 + 10 M^{2} + 2 M \sin T - 11 M^{2} \sin^{2} T]}{4(1 + M \sin T)^{2}}.
\end{equation}
From Equations (\ref{eq48}) and (\ref{eq50}), we observe that the positive cosmological
constant is a decreasing function of time and  approaches a small value in the present epoch. \\
{\bf Some Physical aspects of the Models:} \\
The expansion$(\theta)$ and the shear$(\sigma)$ in the model (\ref{eq44}) are given by
\begin{equation}
\label{eq51} 
 \theta = \frac{2\sqrt{2} ~ M \cos T}{(1 + M \sin T)} 
\end{equation}
\begin{equation}
\label{eq52}
\sigma = \frac{1}{\sqrt{3}} ~ \frac{M \cos T}{(1 + M \sin T)} 
\end{equation}
When $T \rightarrow 0$ then $\theta \rightarrow 2\sqrt{2} ~ M$ and when $T \rightarrow \frac{\pi}{2}$
then $\theta \rightarrow 0$. Thus the expansion in the model starts at $T = 0$ and it stops at
$T = \frac{\pi}{2}$. Since $lim_{T \rightarrow \infty} \frac{\sigma}{\theta} \neq 0$, hence the 
models do not approach isotropy for large values of $T$. The model has singularity at $T = 0$ 
which is real physical singularity.
\section{Conclusions} 
We have obtained a new class of Bianchi type IX anisotropic cosmological
models with a viscous fluid as the source of matter. Generally, the models are
expanding, shearing and non-rotating. In all these models, we observe that they do 
not approach isotropy for large values of time $T$. \\
The cosmological constant in all models given in sections  3.1 and  3.2 are decreasing 
function of time and they all approach a small positive value as time increases (i.e., the
present epoch). The values of cosmological ``constant'' for these models are
found to be small and positive which are supported by the results from recent
supernova observations recently obtained by the High - z Supernova Team and 
Supernova Cosmological Project ( Garnavich {\it et al.}, 1998 ; Perlmutter 
{\it et al.}, 1997, 1998, 1999; Riess {\it et al.}, 1998; Schmidt {\it et al.}
, 1998. Thus, with our approach, we obtain a physically relevant decay 
law for the cosmological constant unlike other investigators where {\it adhoc} 
assumption for the variation  were used to arrive at a mathematical expressions for the 
decaying vacuum energy. \\   
\section*{Acknowledgements} 
\noindent
One of the authors (A. Pradhan) thanks to the Inter-University Centre for Astronomy and 
Astrophysics, India for providing  facility under Associateship Programmes where part of 
work was carried out. Authors would also like to thank Professor Raj Bali for helpful 
discussions.  \\
\newline
\newline

\end{article}
\end{document}